\documentstyle[aps,epsfig]{revtex}
\begin{document}
\draft
\preprint{gr-qc/0302170}
\title{\large
{\bf Black Hole Entropy from a  Highly Excited Elementary String}
\footnote{
Talk at the Workshop on {\em Field Theoretic Aspects of Gravity II,} 
October 2-9, 2001, Ooty, India}}
\author{Romesh K. Kaul 
\footnote{email: kaul@imsc.res.in}}
\address{\it The Institute of Mathematical Sciences, Taramani, Chennai
600 113,
India.}

\maketitle

~
\begin{abstract}   
Suggested correspondence between a black hole and a highly excited 
elementary string is explored. Black hole entropy is calculated by
computing the density of states for an open excited string.  We identify the
square root of  oscillator number of the excited string with Rindler
energy  of black hole to obtain 
an entropy  formula which, not only agrees at the leading order with the
Bekenstein-Hawking entropy, but also reproduces the logarithmic 
correction obtained for black hole entropy in the quantum geometry 
framework. This provides an additional supporting evidence for
correspondence between black holes and strings.
\end{abstract}
          
~

It is now more than a decade since 't Hooft suggested a complementarity
between black holes and strings\cite{tH}. 
Black hole horizon is governed by some conformal operator algebra on a
two-dimensional surface. So is a string.
These two may be equally fundamental pictures, related by a
complementarity. As emphasized by 't Hooft, it may be possible to provide 
a black hole interpretation of strings. Also conversely,
black holes will  have a  string representation.
This idea is  substantially
further developed  by Susskind's suggestion that the spectrum of a
Schwarzschild black hole is described by the states of a highly excited
(uncharged) string at Hagedorn temperature\cite{Su}. This allows
a determination of the black hole entropy by counting  the number of
states of an excited string. Thus we have  a statistical interpretation 
of the black hole entropy. More evidence of this correspondence has been 
subsequently presented in refs.\cite{Ha,Ho,DV,Hal} and many others, 
where a variety of 
cases of black holes, non-rotating and rotating, uncharged and charged, 
in different dimensions are  studied from this 
perspective. In particular,  Bekenstein-Hawking 
entropy of a Schwarzschild black hole in four and also arbitrary dimensions 
is reproduced from the density of states of an excited string. 

The correspondence principle for the two spectra, of a black hole and
a highly excited elementary string, may be understood as follows.
As the string coupling $g_{st}$ ($g_{st}^2 = (\ell_P/\ell_S)^{d-2}$
in $d$ dimensions,  $\ell_P$ is the Planck length  and $\ell_S$ is the 
string length scale) increases, the Compton wave length of a high mass and
low angular momentum string state shrinks to less than its Schwarzschild
radius to become a black hole. On the other hand, as the coupling is
reduced, the  black hole  eventually becomes smaller than the string
size. The metric near the horizon then loses its meaning and instead of
being a black hole such a system is better described as a string state.   
But at some intermediate size, when the black hole size and  string size
are equal, either description is admissible. This would imply a
one-to-one correspondence between the spectra of  black holes and 
strings\cite{Su}.

At first this correspondence between the two sets of states appears to be
beset with a difficulty. As functions of mass, there is an apparent 
striking difference between the black hole density of states and the 
density of string states. The  latter  in any dimensions 
grows exponentially in the first power of mass $M$ of the excited 
string\cite{FV}. This implies string 
entropy as linearly proportional to mass $M$, $S_{st} \sim$
$\sqrt{\alpha^\prime} ~M$$ \sim \sqrt{N}$, where $\alpha^\prime$ is 
the inverse of the string tension and  $N$ is oscillator occupation
number of the excited string state.  In contrast, the density of 
states of a black
hole grows exponentially with the second power of mass in
four dimensions (and in general as $M^{(d-2)/(d-3)}$ in $d$ dimensions).
The corresponding entropy then is proportional to square of the
mass in four dimensions (or to $M^{(d-2)/(d-3)}$ in arbitrary $d$
dimensions). As suggested by Susskind\cite{Su}, this apparent discrepancy
can be cured by a proper identification that takes in to account 
a large mass-renormalization, a gravitational redshift. To do this use 
is made of the fact that the near horizon geometry of a Schwarzschild 
black hole is a Rindler space with a dimensionless time $\tau_R$ and 
a dimensionless energy $E_R$. The Rindler mass and ADM mass of a black 
hole are related by a huge redshift between the stretched horizon and 
asymptotic infinity. In $d$  dimensions $ (d \ge 4 )$ the dimensionless 
Rindler energy associated with a Schwarzschild black hole is given 
by\cite{Su,Ha}  
\begin{equation}
E_R ~=~\left(\frac{2}{d-2}\right)  M^{(d-2)/(d-3)} 
\left( \frac{16\pi G}{(d-2) \;~A_{d-2}} \right)^{1/(d-3)} 
=~ \left(\frac{2}{d-2}\right)  M~ r_{BH},
\end{equation} 
where $A_{d-2}$ is the area of a unit sphere of  $d-2$ dimensions,
$G$ is the Newton's constant and
$r_{BH}$ is the Schwarzschild radius associated with mass $M$.
In terms of  Rindler energy, the  horizon `area' $A_H$  and 
Bekenstein-Hawking entropy $S_{BH}$ of a Schwarzschild black hole 
in any arbitrary dimensions is given by
\begin{equation}  
A_H~ =~ 8\pi G E_R, ~~~~~~~~~~~~~ S_{BH}~=~ 2\pi E_R~. 
\end{equation}  
That is, black hole density of states grows exponentially 
with Rindler energy in any dimensions. The 
{\it string}$\Leftrightarrow${\it black-hole} correspondence is then obtained
when  Rindler energy $E_R$ is identified with the square root of
the oscillator number $\sqrt{N}$ of the highly excited string in
{\it any dimension}. In particular, in four dimensions, ~$ N ~\sim~
E_R^2 ~ \sim G^2 M^4$.  This
identification, then gives the same leading order
expression for the density of black hole states and that for the string
states allowing  equality of  Bekenstein-Hawking
entropy of a black hole with that of an excited string in the leading
order. Further support for
this correspondence is the consequent identification of  Hawking 
temperature of the black hole with   red-shifted Hagedorn temperature 
associated 
with the excited string.  Equivalently,
the black hole size $r_{BH}$ will get identified with the string
length scale $l_S $.

If the correspondence ~{\it black-hole}$\Leftrightarrow${\it  string},
with the identification of Rindler energy ($E_R$) with the square root of
the oscillator occupation number ($\sqrt{N}$) of  highly excited string, 
is strictly true, all other
features of black holes must get reflected in the string description
too. In this context, $ln(area)$ corrections  to the Bekenstein-Hawking
entropy, first discovered in  the quantum geometry framework,
is of interest. In this framework, boundary degrees of
freedom of a black hole in four dimensional gravity are described by 
an $SU(2)$ Chern-Simons theory on the horizon, with coupling $k$ 
proportional to the horizon area. The dimensionality of boundary 
Hilbert space can thus be readily computed by counting the conformal 
blocks of $SU(2)_k$ conformal field theory on a two-sphere (spatial slice 
of the horizon) with a number of punctures carrying $SU(2)$ spin
representations on them\cite{KM1}.
Then for large horizon area $A_H$, 
the black hole entropy $S_{bh}$ has been found to be\cite{KM2,DKM}   
\begin{equation}{\label3}
  S_{bh} ~=~ \frac{A_H}{4G} ~-~ \frac{3}{2} \; ln\left(\frac{A_H}{4G}\right)
~+~\cdots \cdots\;
\end{equation}
There have also been other subsequent derivations of this  $ln(area)$  
correction with the same  coefficient $-3/2$ \cite{Ca,GKS,BS,G}. It 
appears to be
universal in the sense that it obtains for a variety of black holes and
also in different dimensions. 

Thus, if ~{\it black-hole}$\Leftrightarrow${\it  string}~
correspondence is indeed true, the same correction as in eqn.(\ref{3})
should be reflected in the
entropy of a highly excited string also. Over years, the counting  of
string states has been done in many places. In the following, we
shall recalculate {\it the level density of a highly excited open string
beyond the leading order carefully taking into account  contribution
of the zero modes.} It will be  
demonstrated  that the string entropy does indeed receive a correction
$-3/2 ~ln \sqrt{N} $ beyond the leading value proportional to $\sqrt{N}$. 
This may then be taken as an additional evidence in support of
the ~{\it  black-hole}$\Leftrightarrow ${\it string}~ correspondence. 

An open string moving in a $d$ dimensional space-time is described by
$d$ two-dimensional fields $X^\mu(\sigma,\tau),$ $\mu = 0, 1, 2,
....(d-1)$ with $0\le \sigma \le \pi ,$ satisfying the string equation
$\partial^2 X^\mu (\sigma, \tau)  - \partial^2X^\mu (\sigma, \tau)  =  0$
subject to  boundary conditions reflecting no flow of
momentum from the string ends:  $\partial_{\sigma} X^\mu (\sigma,\tau) =0$
at $\sigma = 0$ and $\sigma = \pi .$
The string equation is satisfied by
\[
X^\mu ~ =~ x^\mu ~+~ \ell^2_S ~p^\mu~ \tau ~+~ i\ell_S \sum_{n\ne0}
\frac{1}{n} ~\alpha^\mu_n ~ e^{-in\tau} ~cosn\sigma
\]
where~ $\ell_S = \sqrt{2\alpha^\prime} = 1/\sqrt{\pi T}$ ~is the
fundamental string length (our units are $\hbar = c = 1),$ $T$ is
string tension. The center of mass coordinate is $x^\mu $ and string
momentum is $p^\mu$. The commutation relations satisfied by the various
operators are
\[
[~x^\mu ~, ~p^\mu ~] ~=~ i\eta ^{\mu \nu}, ~~~~~~~[\alpha^\mu_m ~,
~\alpha^\nu_n~] ~=~ m ~\delta_{m+n, 0} ~\eta^{\mu \nu},
\]
where $\eta ^{\mu \nu}$ is the flat metric in $d$ dimensional
space-time. This theory has a reparametrization invariance
$(\sigma, \tau) \rightarrow (\sigma^\prime, \tau^\prime)$ 
generated by Virasoro constraints: $(\partial_\tau X^\mu \pm 
\partial_\sigma X^\mu)^2 =0. $~ This  may be fixed by
a gauge choice, in particular by light-cone gauge introducing a preferred
longitudinal direction in space:
$X^{+}  = x^{+} + \ell^2_S~ p^{+} ~\tau$,
where the light-cone coordinates are~ $ X^\pm = (X^0 \pm 
X^{d-1})/\sqrt{2}$. In this gauge all the oscillators $\alpha^{+}_n = 0 $
~$(n\ne0)$ and also the coordinates $X^- (\sigma, \tau)$ and hence
$p^-$ and $\alpha^{-}_n$ are not independent.  Only independent
variables are transverse coordinates, momenta and oscillators:
$x^i$, ~$p^i$, ~$\alpha^i_n$~ $(n \ne0)$ with $i~=~ 1,~2,~3, ....d -2$.
The $\tau$-translation generating Hamiltonian is given by
\begin{equation}
H~=~\ell^2_S~ p^+p^- ~=~ \frac{1}{2}~\ell_S^2~ (p^i)^2 
 ~+~ {\cal N} ~-~ a 
\end{equation}
where ~ $a = (d-2)/24$ ~is the normal ordering constant and 
${\cal N} = \sum^{d-2}_{i=1} \sum^{\infty}_{m=1}
\alpha^i_{-m}~\alpha^i_m$~ is the oscillator number operator whose eigenvalues 
are the occupation number of the state, $ {\cal N} ~|~{\psi_N}>
=N ~|~{\psi_N}>$. 

It is convenient to introduce the standard
oscillators:
\[ {a^i_m}^\dagger~=~ \frac{1}{\sqrt{m}}\alpha^i_{-m}, ~~~~~~~a^i_m ~=~
{\frac{1}{\sqrt m}} \alpha^i_m   ~~~~~~m >0
\]
which satisfy the standard oscillator commutation relations:
\[
[~ a^i_m, ~{a^j_n}^\dagger~] ~=~ \delta_{mn}~\delta^{ij} .
\]
The occupation number operator in terms of number operators ${\cal N}_m$
for these standard oscillators is
\begin{equation}  
{\cal N} ~=~ \sum^{\infty}_{m=1} m~{\cal N}_m ~\equiv~ \sum^{\infty}_{m=1}
~\sum^{d-2}_{i=1} ~m~{a^i_m}^\dagger a^i_m,
\end{equation}
where ${\cal N}_m $ has the standard oscillator eigenvalues $0,~1,~2,~3.....$.

The mass of an excited string of level $N$  is given by 
$\ell_S^2 M_N^2 = 2(N-a)$. To count the quantum states of such a string, 
we write the partition function  as a function of  a complex
parameter $\tau$ as (we set $\ell_S =1$ in the following):
\begin{equation}
Z(\tau) ~=~ Tr ~e^{2\pi i H \tau} ~=~ Tr ~exp\left[ 2\pi i \tau
\left( \frac{(p^i)^2}{2} ~+~ {\cal N} ~-~ a \right) ~\right].
\end{equation}
Here  $Tr$ represents integration over the transverse string momentum
$p^i$  and trace over 
the oscillator states. That is,
\begin{equation}\label{7}  
Z(\tau) ~=~ \int \left(d^{d-2}p^i\right) e^{\pi i\tau (p^i)^2} tr ~exp \left[ 2\pi
i \tau \left( {\cal N} ~-~ a \right) \right] ~=~ \left( \frac{1}{-i\tau}
 \right)^{(d-2)/2}  e^{-2\pi i \tau a} ~tr ~exp \left[ 2\pi i\tau  
{\cal N} \right], 
\end{equation}
where now $tr$ represents trace over the oscillator states only. Notice that
\[
tr ~ e^{2\pi i \tau {\cal N} } ~=~ \sum^{\infty}_{m=1} e^{2\pi im\tau {\cal N}_m}
~=~ \left(\sum^\infty_{N=1} ~ p(N) e^{2 \pi i N \tau} \right) ^{d-2}
\]
where $p(N)$ is the number of partitions of $N$ in terms of positive integers.
Now
there is a standard formula in number theory:
\[
 f^{-1}(\tau)~ \equiv~ \sum^{\infty}_{N=1}~ p(N) ~e^{2 \pi iN \tau} ~=~ \prod^{\infty}_{n=1} \left(
1~-~ e^{2 \pi in \tau} \right)^{-1}.
\]
The function $f(\tau)$  is related to  Dedekind eta function as: $\eta (\tau) = exp(i \pi
\tau /12) ~f(\tau )$. Thus the partition function is
\[
Z(\tau ) ~=~ \left( \frac{1}{-i \tau} \right)^{(d-2)/2}
~\frac{1}{\left[\eta (\tau)\right]^{d-2}} ~.
\]
Next note Dedekind eta function has the property: $\eta (-1/\tau ) =
(-i \tau)^{1/2} ~\eta (\tau ). $ Using this, the  partition function can be 
written as  
\begin{equation}\label{8}
Z(\tau ) ~=~  \frac{1}{\left[ \eta (-1/\tau) \right]^{d-2}}~=~ \frac{e^{2\pi
ia/\tau}} 
{\left[f(-{1}/{\tau})\right]^{d-2}}
\end{equation}

To find the  density  ~$d(N)$~ of  string states with occupation
number $N$,~ we write
\begin{equation}\label{9}
Z(\tau ) ~=~ \sum^{\infty}_{N=0} d(N) ~ e^{2 \pi i(N ~-~ a) \tau} ~ .
\end{equation}
Equating the two expressions in eqns.(\ref{8}) and (\ref{9}) and inverting
for the level density, we have
\begin{equation} 
d(N) ~=~ \int d\tau ~~\frac{exp\left[-2 \pi i \left(~(N~-~a) \tau ~-~
(a/\tau)\right)\right]}{\left[f(-{1}/{\tau})\right]^{d-2}} ~.
\end{equation}
For large $N$, this can now be approximately evaluated  by the saddle 
point method. The saddle point  is at $\tau_0  = i \sqrt{a/(N-a)}$. 
Expanding around this point ~$\tau= \tau_0 + u$ and  performing the  Gaussian
integration  over $u$  yields the level density as
\[
d(N) ~\simeq~  \left( \frac{a^{1/2}}{2(N~-~a)^{3/2}}
\right)^{1/2} exp \left( 4 \pi {\sqrt{a (N~-~a)}} \right),
\]
where $f(-1/\tau_0) \rightarrow 1$ for large $N$ has been used.
Finally  density of string states  for large occupation number $N$ can 
be  asymptotically written as:
\begin{equation}\label{11}  
d(N) ~ \simeq ~C  ~\frac{a~exp\left( 4\pi \sqrt{aN}\right)}{(aN)^{3/4}} ~~+~~ \cdots
\cdots~, 
\end{equation} 
where  $C$ is an $N$ independent irrelevant constant and
$a = (d-2)/24$. This is the same formula as that obtained by
Carlip for density of states with eigenvalues of Virasoro operator 
$L_0$ as $\Delta = N - a$ ~in a general rational conformal field 
theory of central charge $c=24a$\cite{Ca}. 
The asymptotic level density of string states has been calculated  
in many places, the earliest computation for a string in $d=26$ 
dimensions was done in refs.\cite{FV} where the rapidly growing
exponential dependence on $\sqrt{N}$ was correctly obtained.
However,  there is now  an additional factor of $N^{-3/4}$. 
This introduces  a logarithmic correction to  the  entropy  $S_{st}$ of
a highly excited string: 
\begin{equation}
S_{st} ~=~ ln~d(N) ~\simeq~ 4\pi ~\sqrt{aN} ~-~ \frac{3}{2}~ln \sqrt{aN} ~+
~lna ~ - 
~\cdots  \cdots
\end{equation}

Finally  following Susskind we identify Rindler energy of the black hole with 
$\sqrt{N}$ as $E_R = 2~\sqrt{ a N}$,  and rewrite  the entropy of 
a highly excited string as 
\begin{equation}
S_{st} ~=~ 2\pi E_R ~-~ \frac{3}{2} ~ln E_R ~-~ \cdots \cdots
\end{equation}
Clearly  this entropy has  same logarithmic correction  beyond the
Bekenstein-Hawking area law as  that obtained
for the black hole entropy in  quantum geometry framework\cite{KM2}. 
This  correction   exits not only for excited strings 
describing Schwarzschild black holes in any arbitrary dimensions
as above, but also all other cases discussed in 
refs.\cite{Ha,Ho,Hal}.
We emphasize that this thus provides an additional evidence in favour of the 
{\it excited-string}$\Leftrightarrow${\it black-hole} correspondence. 

In the derivation of level density above, it is important to take
account of the zero modes carefully. This has been done by including
the integration over $p^i$ in the partition function in eqn.(\ref{7}).
If this were not included, then non-exponential  part  of level density
formula (\ref{11}) above would have changed from dimension independent
factor $N^{-3/4}$ to $N^{-(d+1)/4}$ as has been found  
in some of the early calculations of level density 
(for example, see ref.\cite{GSW} for the case of $d=26$). 

Our discussion here has been for generic black holes, say, Schwarzschild
black holes in any dimensions. Though level density  
above has been calculated for an open string, the
asymptotic formula (\ref{11}) is valid in general in any string
theory.  In particular, this also obtains for the level density
of BPS elementary string states of the superstring
theories. For extremal black holes of these  superstring theories for 
which no mass renormalization takes place, the  correspondence  between
strings and black holes sharing the same macroscopic quantum numbers
may be applied  directly. This allows a  
counting of  weakly coupled BPS string states which can be directly 
related to degeneracy of these extremal black holes reproducing 
the Bekeintein-Hawking entropy in the leading order\cite{S,SV}.
The correction beyond  Bekenstein-Hawking  entropy  obtained here
holds for  the Bogomol'nyi saturated elementary string states
too\cite{Ca}.

Like black holes, entropy of de Sitter space can also be given a string
interpretation\cite{Ha1}. Here also geometry near the cosmological horizon is
given by a Rindler space. Thus  de Sitter space  may well be described
by a string on  the stretched horizon through the same identification  of 
Rindler energy with  square root of the oscillator number. The entropy 
so calculated, beyond the usual area law would also have the same
logarithmic correction as  discussed above.

~

{\bf Acknowledgement:} Discussions  with  S. Kalyana Rama  are gratefully
acknowledged.


\begin{references}
 
\bibitem{tH}
G. 't Hooft, Nucl. Phys. {\bf B335} (1990) 138.

\bibitem{Su}
L. Susskind, {\em Some Speculations about Black Hole Entropy in
String Theories,} hep-th/9309145.

\bibitem{Ha}
E. Halyo, A. Rajaraman and L. Susskind, Phys. Letts. {\bf B392}
(1997) 319;
E. Halyo, B. Kol, A. Rajaraman and L. Susskind, Phys. Letts. {\bf B401}
(1997) 15.
                                                    

\bibitem{Ho}
G.T. Horowitz and J. Polchinski, Phys. Rev. {\bf D55} (1997)
6189. 

\bibitem{DV}
T. Damour and G. Veneziano, Nucl. Phys. {\bf B568} (2000) 93.

\bibitem{Hal}
E. Halyo, Jour. High Energy Physics {\bf 0112} (2001) 005.  

\bibitem{FV}
S. Fubini and G. Veneziano, Nuovo Cim. {\bf A64} (1969)811;
K. Huang and S. Weinberg, Phys. Rev. Letts. {\bf 25} (1970) 895.

\bibitem{KM1}
R. K. Kaul and P. Majumdar,
Phys. Lett. {\bf B439} (1998)267.
 
\bibitem{KM2}
R. K. Kaul and P. Majumdar,
Phys. Rev. Lett. {\bf 84} (2000) 5255.
 
\bibitem{DKM}
S. Das, R. K. Kaul, and P. Majumdar,
Phys. Rev. {\bf D63} (2001) 044019.
                   

\bibitem{Ca}
S. Carlip,
Class. Quant. Grav. {\bf 17} (2000) 4175.
 
\bibitem{GKS}
T. R. Govindarajan, R. K. Kaul, and V. Suneeta,
Class. Quant. Grav. {\bf 18} (2001) 2877.
 
\bibitem{BS}
D. Birmingham and S. Sen,
Phys. Rev. {\bf D63} (2001) 047501;
K. S. Gupta and S. Sen,
Phys. Lett. {\bf B526} (2002) 121;
K. S. Gupta, {\em Near Horizon Conformal Structure and Entropy of 
Schwarzschild Black Holes,} hep-th/0204137.
           
\bibitem{G}
G. Gour,
Phys. Rev. {\bf D66} (2002) 104022.

\bibitem{GSW} 
M.B. Green, J.H. Schwarz and  E. Witten, {\em Superstring Theory, vol 1,
page 118},
Cambridge University Press, 1987.

\bibitem{S}
A. Sen, Mod. Phys. Lett {\bf A10} (1995) 2081.
 
\bibitem{SV}
A. Strominger and C. Vafa, Phys. Letts. {\bf B379} (1996) 99.
                                                 
\bibitem{Ha1}
E. Halyo, {\em De Sitter Entropy and Strings}, hep-th/0107169.

\end{references}
\end{document}